\documentclass{raa}           
\usepackage{graphicx,times}
\usepackage{natbib}
\usepackage{amssymb,amsmath}
\bibpunct{(}{)}{;}{a}{}{,}

\usepackage[T1]{fontenc}
\usepackage{tikz}
\usetikzlibrary{shapes.geometric}
\usetikzlibrary{arrows,calc,positioning}

\tikzstyle{intg}=[draw,minimum size=3em,text centered,text width=6.cm]
\tikzstyle{decision} = [diamond, draw, thick,
text width=4.5em, text badly centered, inner sep=1pt]

\usepackage[a4paper=true,pagebackref=true,driverfallback=dvipdfm]{hyperref}
\hypersetup{pdftitle = The title of my PDF, pdfauthor = My name, pdfsubject= The subject, pdfkeywords = keyword1 keyword2 keyword3} 
\hypersetup{colorlinks = true, linkcolor = green, anchorcolor = red, citecolor = blue, filecolor = red, urlcolor = red}

\begin{document}

\title{Development of image motion compensation system for 1.3 m telescope at Vainu Bappu Observatory
}

 \volnopage{ {\bf 20XX} Vol.\ {\bf X} No. {\bf XX}, 000--000}
   \setcounter{page}{1}

   \author{Sreekanth Reddy V\inst{1}$^*$, Ravinder Kumar Banyal\inst{1}, Sridharan R
      \inst{1},   P U Kamath\inst{1}, Aishwarya Selvaraj\inst{1}
   }

   \institute{Indian Institute of Astrophysics, Bangalore, 560034, India; {\it $^*$sreekanth@iiap.res.in}\\
\vs \no
   {\small Received 20XX Month Day; accepted 20XX Month Day}
}

\abstract{We developed a  tip-tilt system to compensate the turbulence induced image motion for 1.3 m telescope at Vainu Bappu Observatory, Kavalur. The instrument is designed to operate at visible wavelength band (480-700 nm) with a  field of view $1^{\prime}\times1^{\prime}$. 
The tilt corrected images have shown up to $\approx$ 57\% improvement in image resolution and a corresponding peak intensity increase by a factor of $\approx$~2.8.  A closed-loop correction bandwidth of $\approx$~26~Hz has been achieved with on-sky tests and the \emph{root mean square} motion of the star image has been reduced by a factor of $\sim$ 14. These results are consistent with theoretical and numerical predictions of wave-front aberrations caused by atmospheric turbulence and image quality improvement expected from a real-time control system. In this paper, we present the  details of the instrument design, laboratory calibration studies and quantify its performance on the telescope.
}
   \authorrunning{V S Reddy et al. }            
   \titlerunning{Image motion compensation system}  
   \maketitle
%
\section{introduction}\label{sec:intro}
Soon after the telescope was discovered, it was realized that atmospheric turbulence poses a major hindrance to the performance of the ground-based telescopes. The effects of atmospheric turbulence on astronomical observations  was investigated by \cite{fri65,Fried66} by relating the statistics of wave distortion to optical resolution. The extensive research on the effect of atmospheric studies shows that the imaging through turbulence is limited by atmospheric seeing $\lambda/r_0$, irrespective of optical system resolution $\lambda/D$, where $r_0$ is Fried parameter and $D$ is the aperture diameter.  

The turbulence induced wave-front distortions are often spread over different spatial scales. The study by \cite{noll76} expressed these distortions using Zernike polynomials representing varying degree of aberrations. These polynomials are widely used to evaluate optical system performance. The same was used by \cite{noll76} to estimate the effect of atmospheric turbulence on wave-front. From these investigations, it became clear that  the effect of wave-front distortions scale with the size of the system aperture. Larger the telescope, stronger the deleterious effect of the higher-order distortions. The image motion, being the lowest order aberration, is  primarily caused by the global tilt in the wave-front -sometime also called the angle of arrival fluctuations. The contribution of the lowest order distortions is about 87\% of the wave-front phase variance (\citealt{fri65, dainty98}). By eliminating the image motion using a real-time tip-tilt system, the image resolution can be significantly improved -at least for small aperture diameters.

The optical wave-front correction with adaptive optics (AO) system was first used by the US Navy for defence purpose (\citealt{greenwood77}). Its enormous potential was recognized in the field of medical and astronomical use. One of the early realizations of  astronomical AO systems, the \emph{COME-ON} prototype system, was on Observatoire de Haute-Provence (\citealt{rousset90}). Modern telescopes equipped with AO are highly productive and serving the astronomical community better. 

The design and development of an AO system should consider the atmospheric characteristics of the telescopic site.  Apart from telescope size, the Fried parameter(\citealt{fri66}), isoplanatic angle (\citealt{hubbard79}) and coherence time (\citealt{davis96,kel07}) are the key parameters that can guide the design of AO system. The Fried parameter determines the minimum spatial sampling of the wave-front for sensing and correction. The coherence time helps to determine the optimal loop frequency for close loop AO operation. For high-speed operation, the target star should have enough photons to adequately sample the wave-front -both temporally and spatially. This limits the operation of natural guide star AO systems to a small number of target stars that are relatively brighter (\citealt{wizinowich2000}). Therefore, to increase the overall sky coverage and improve the effective AO correction over large angular range, a single or multiple artificial  laser guide stars are often deployed (\citealt{max97}).

The knowledge of isoplanatic angle is essential in natural guide star systems. In the case of fainter science target, the AO system uses the nearby bright target as a guide star, preferably within an isoplanatic angle. The proximity of the science target to the guide star defines the active correction of the wave-front. Typically, this parameter will range from 2$^{\prime\prime}$-6$^{\prime\prime}$(\citealt{eaton85,sarazin02}).

\begin{figure}
\centering
\includegraphics[height=9.5cm]{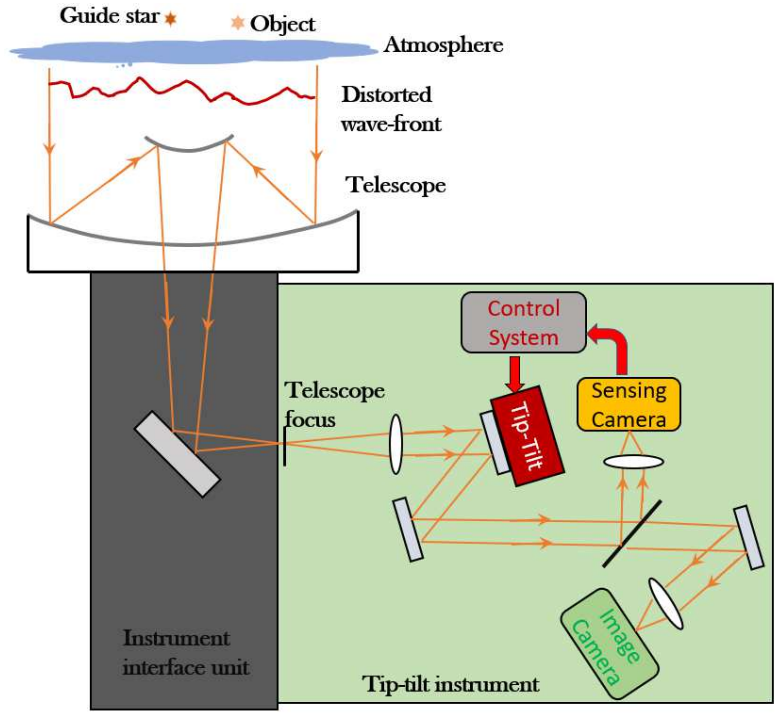}
\caption{ \label{fig:AOLayout} Conceptual layout of a natural guide star tip-tilt correction system. The telescope primary mirror collects the light from a target field. The light beam is directed to tip-tilt instrument. The light is divided between the sensing arm and the imaging arm of the instrument. The image motion is corrected in common path where the tip-tilt stage is placed. The corrected wave-front can be directed to imaging camera or a science instrument.}
\end{figure}

Conventionally, AO systems initially correct the slope (global tilt with  large amplitude) of the wave-front before they correct high-order distortions (high spatial frequency terms with smaller amplitudes). In the former case, a rapidly moving 2-axis steering mirror is used for correcting the wave-front slope variations. Such a system is conceptually simple to develop and  cost effective. The layout of a natural guide star tip-tilt correction system is shown in Figure \ref{fig:AOLayout}. This system has a tip-tilt stage with a mirror mounted on it. In response to changing conditions, the tip-tilt controller constantly steers the actuators to keep stellar beam locked to a fixed reference position in the image plane. A number of tip-tilt systems have been developed in the past (\citealt{racine89, glindemann1997charm}). A tip-tilt system at Calar Alto 3.5~m telescope, e.g., has shown image motion reduction from $\sim$ $\pm$0.4$^{\prime\prime}$ to $\sim$ 0.03$^{\prime\prime}$  with 30-100~Hz loop frequency  (\citealt{glindemann1997charm}). This alone could yield a significant improvement in image resolution. A proper exposure time of a tip-tilt sensor is crucial to optimize the performance of the instrument (\citealt{mar87}). For example,  \cite{glindemann97} has shown that to track/correct an image motion with 5-10 Hz bandwidth, a loop frequency of 50 -100 Hz is required. \cite{close94} have developed a Cassegrain secondary tip-tilt AO. They have reported three-fold  image motion reduction at 72~Hz loop frequency. Furthermore, \cite{golimowski92} have used image motion compensation system for high resolution stellar coronography. The instrument is reported to have achieved a resolution gain by a factor of 2.2. It enabled the observations of a two magnitudes fainter objects than what was achieved without the image stabilization.

The Indian Institute of Astrophysics has initiated  a long-term project to develop AO systems for its observatory telescopes. We plan to achieve this in three phases: first, estimating the  turbulence parameters at the telescope site, second, the development of tip-tilt image motion compensation system and third, the design and development of higher order AO system. Atmospheric turbulence parameters, namely, the Fried parameter, isokinetic angle and coherence time have been measured in the initial phase (\citealt{VSReddy2019}).  In this paper we discuss the design and development of a tip-tilt system for 1.3~m telescope at Vainu Bappu Observatory at Kavalur. From here on the paper is organized as follows. In Section \ref{sec:develop} we present the details of opto-mechanical design of the instrument followed by the development of control software and tip-tilt calibration methodology. Laboratory tests and characterization of the instrument are discussed in Section \ref{sec:labtest}. The on-sky performance of the instrument is described in Section \ref{sec:onsky}. Finally, in Section \ref{sec:discuss}, we summarized our results.

\section{Opto-mechanical design}\label{sec:develop}
The tip-tilt system was designed for 1.3 m telescope, Vainu Bappu Observatory located at $78^\circ 50^\prime E$ and $12^\circ 34^\prime N$. The telescope was commissioned in the year 2014. It is a Ritchey-Chretien model telescope with primary and secondary being hyperbolic mirrors. During the routine observations  telescope resolution is seeing-limited. A tip-tilt instrument followed by a higher order AO system is envisioned to transform the seeing-limited resolution of the telescope to near diffraction limited case.

\subsection{Opto-mechanical design}
An optical system of the instrument was designed in ZEMAX ray tracing software. Several design parameters, such as effective focal length, overall weight and dimensions of the instrument were considered. The pixel scale at telescope Cassegrain focus is 0.26$^{\prime\prime}$/pixel (pixel size =  13 $\mu$m ) while the diffraction limited resolution is 0.106$^{\prime\prime}$ at 630 nm wavelength. In the optical design of the instrument, the pixel scale was reduced to half of the diffraction limited resolution. This improved the sensitivity of the telescope to measure the image motion. The instrument weighs $\approx$ 28~kg and has dimensions $120\times60\times30$ cm$^{3}$. It was installed on the West port of the telescope as shown in the top panel of Figure~2. A solid model  designed in $\it AutoCAD$ is shown in bottom panel of Figure~2.  Some key  specifications of the telescope and the tip-tilt instrument are listed in Table~\ref{tab:Telescope}.

\begin{figure}
\centering
\includegraphics[height=11.5cm]{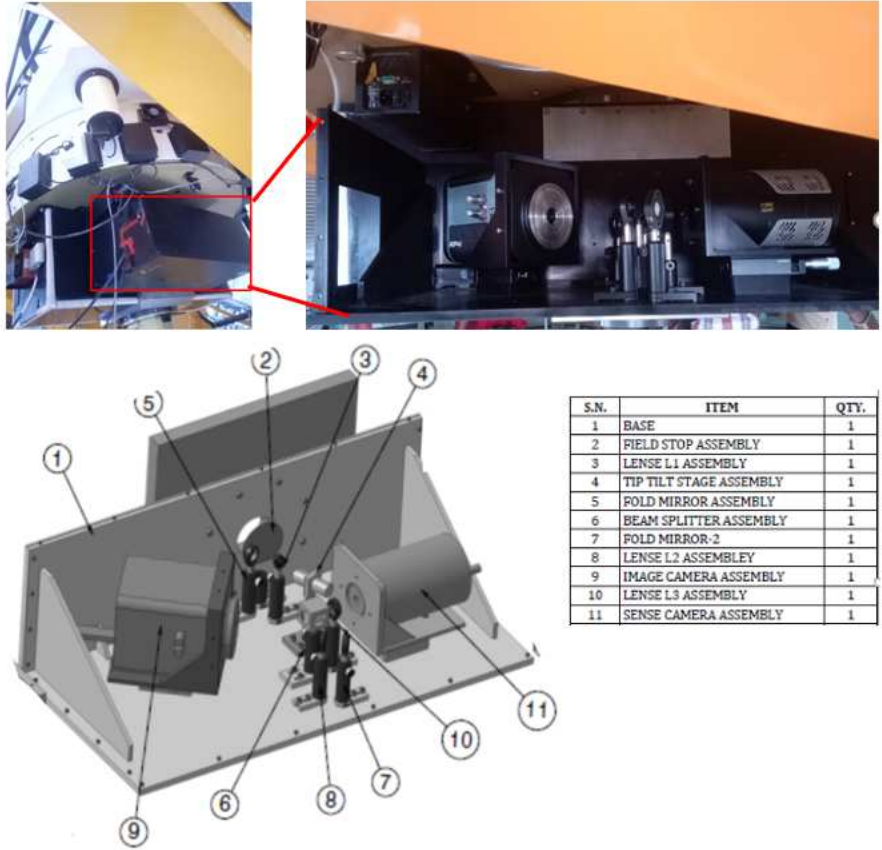}
\caption{ \label{fig:instrumentMech} The tip-tilt instrument mounted on the West port of Cassegrain focus of 1.3 m JCB telescope at VBO (top). A CAD model showing the mechanical layout and system components (bottom).}
\end{figure}

\begin{table}
    \bc
	\centering
	\caption{Key specifications of tip-tilt instrument. Telescope and instrument layout  is shown in Figure \ref{fig:AOLayout}. F/$_{\#}$: F-ratio, PS: Pixel Scale.}
	\label{tab:Telescope}
	\begin{tabular}{lccr} 
		\hline\noalign{\smallskip}
		Property & Value \\
		\hline\noalign{\smallskip}
		Wavelength range& 480-700 nm\\
    	wave-front sense plane FOV2 & 40$^{\prime\prime}$\\
    	Image plane FOV1 & 1$^{\prime}$\\
    	Telescope focus F/$_{\#}$, PS & 8, 0.26$^{\prime\prime}$\\
    	Sensor camera F/$_{\#}$, PS & 15.7, 0.06$^{\prime\prime}$\\
    	Imaging camera F/$_{\#}$, PS & 22.4, 0.08$^{\prime\prime}$\\
		\noalign{\smallskip}\hline
	\end{tabular}
	\ec
\end{table}

\subsection{Sub-components of the instrument}
A brief description of essential sub-components of tip-tilt instrument is as follows. 
\begin{itemize}
    \item[$\bullet$] \textbf{Tip-tilt stage}: We used a fast steering tip-tilt stage  from  \textit{Physik Instrumente}\footnote{https://www.physikinstrumente.com/} (Model: S-330).  It is a Piezo-based actuator system with two orthogonal axis. Each of the axis has two actuators that work on the opposite polarity of an applied voltage. These axis are able to deflect the light beam with 0.5 $\mu$rad resolution over a 10~mrad tilt angle. The tip-tilt stage has a one-inch diameter platform to mount a mirror that steers the light beam projected on to it. The stage is driven by  \textit{Physik Instrumente} E-517 control system that has proportional, integral and differential (PID) internal voltage controller to provide a position accuracy of 0.5 $\mu$rad in a close-loop operation.
    
    \item[$\bullet$] \textbf{Sensing camera}: The performance of the tip-tilt instrument depends on its ability to sense the image motion and apply the necessary correction. The sensing camera should be able to operate at high frame rate to record the random drift in the stellar image. We used  \textit{Andor Neo-sCMOS} 2560$\times$2160 pixel, high-speed camera\footnote{https://andor.oxinst.com}. For exposure time of 1 ms and frame size of $128\times128$ pixels, we could able to achieve frame rate of $\sim$\ 300 using control software developed on LabView platform. With 6.5 $\mu$m pixel size and the magnification achieved by the relay optics, the sensing camera is able to sample the sky with 0.06$^{\prime\prime}$/pixel.
    \item[$\bullet$] \textbf{Imaging camera}: We used \textit{Princeton Instruments} ProEm eXcelon ($1024\times1024$, pixel size = $13 \mu$m) EMCCD\footnote{https://www.princetoninstruments.com/} for tilt corrected observations. The CCD is used in continuous exposure mode with simultaneous read out of the data.  This is a frame transfer operation. In this mode, the data in active area is vertically shifted to a masked area. This operation takes few microseconds (0.8 $\mu$sec) and enables the active area to be available for next exposure. The full frame of the CCD cover $\approx 1^{\prime} \times 1^{\prime}$ on-sky field of view (FoV) with plate scale of $0.08^{\prime\prime}$/pixel. 
\end{itemize}

\subsection{Estimation of image motion}
The sensing camera is operated in high-speed mode ($\sim$\ 300 fps) to acquire short exposure images. These images were used to estimate the image motion using a centroid tracking method (\citealt{close94,golimowski92}). The noise in the images affects the accuracy of centroid tracking. The images were bias subtracted.  Although the magnitude of the dark current itself is negligible, there is considerable  bias counts ($\sim$100 counts at 30 K), which contributes error to the centroid estimation. Thus this `dark' subtraction (which includes bias) found to be was essential. The master dark is obtained by taking the median of the dark frames before each experiment. 

 The centroid estimation should be faster to minimize the time delay between the sensing of the motion and the correction.  For this experiment, we have chosen an intensity thresholding centroid technique. In this method, a threshold slightly above the pixel noise level is applied to the image. The method minimizes the noise by assigning zero counts to pixels below the threshold. The resultant image centroid is measured using the weighted average of the intensities as shown in the Equation \ref{eq:eq1}. 
 To measure the improvement in tilt corrected image motion, the root mean square (rms) of the residual centroid motion is estimated using Equation \ref{eq:eq2}:

\begin{center}
\begin{eqnarray}
X_c &=& \frac{\sum x_i I_i }{ \sum I_i}, Y_c = \frac{\sum y_i I_i }{ \sum I_i}, \label {eq:eq1} \\
\sigma_x &=& \sqrt[]{\frac{\sum ({X_c-\bar {X_c}})^2 }{n}}, \sigma_y = \sqrt[]{\frac{\sum ({Y_c-\bar {Y_c}})^2 }{n}} \label{eq:eq2} 
\end{eqnarray}
\end{center}

where $X_c$, $Y_c$ are the estimated centroid of the image, $I_{i,j}$ are pixel intensities, $x_{i,j}$, $y_{i,j}$ are pixel coordinates, $\bar {X_c}$, $\bar {Y_c}$  are mean centroids and $\sigma_x$, $\sigma_y$ are associated standard deviations.

\subsection{Power spectral density}
The power spectral density (PSD) is the measure of energy distributed over a frequency range when the measurements are made within a finite time window (\citealt{welch67}). In this paper, the PSD of the centroids of tilt uncorrected and corrected images was measured. The comparison between the PSD help in determining the correction bandwidth of the tip-tilt instrument. The correction bandwidth is defined as the least frequency at which the ratio of the PSD of the uncorrected and corrected data sets falls to unity, i.e., 0~dB. 
\begin{equation}
	\widehat{x}(f)=\frac{1}{2\pi\sqrt{T}}\int_{0}^{T} e^{-2{\pi}ift}x(t)dt
	\label {eq:eq3}\\
    \end{equation}    
\begin{equation}
	S(f)=|\widehat{x}(f)|^2
	\label {eq:eq4}\\
    \end{equation} 
    The power spectral density $S(f)$ over limited time interval $[0,T]$ of centroid data $x(t)$ is defined in Equation \ref{eq:eq3},\ref{eq:eq4}. Where, $\widehat{x}(f)$ is the Fourier transform of centroid data with temporal frequency $f$.

\subsection{Control Software}
\begin{figure}[!htb]
    \centering
    \begin{tikzpicture}[
      >=latex',
      auto
    ]
      \node [intg] (kp)  {Initialize the system (Start the camera, tip-tilt controller and set all the essential system parameters)};
      \node [intg]  (ki2) [node distance=2.3 cm ,below of = kp] {Camera: Obtain image frame};
      \node [intg]  (ki3) [node distance=1.8 cm ,below of = ki2] {Process Image (Dark subtraction)};
      \node [intg] (ki4) [node distance=1.8 cm,below of=ki3] {Find image centroid (Gravity centroid on intensity threshold)};
       \node [decision] (ki5) [node distance=2.5 cm,below of=ki4] {Closed/Open loop?};
      \node [intg] (ki6) [node distance=2.5 cm,below of=ki5] {Correct for tilts};
      \node [intg] (ki7) [node distance=1.5 cm,below of=ki6] {Tilt corrected (index the image and centroid data)};
      \node [decision] (ki8) [node distance=2.0 cm,below of=ki7] {Stop?};
      \node [intg] (ki9) [node distance=2.2 cm,below of=ki8] {Stop acquisition, correction and save data (images and centroid)};
    \node [intg] (ki10) [node distance=1.5 cm,below of=ki9] {End};
      \draw[->] (kp) -- (ki2);
      \draw[->] (ki2) -- (ki3);
      \draw[->] ($(ki2.south)+(-0.0,-0.3)$) -- ($(ki2.west)+(-1.5,-0.8)$) |- node [near start] {Master loop} (ki2);
      \draw[->] (ki3) -- (ki4);
      \draw[->] (ki4) -- (ki5);
      \draw[->] (ki5.west) -- ($(ki5.west)+(-2.5,0.0)$) |- node [near start] {Open loop} (ki3);
      \draw[->] (ki5.south) -- node [near start] {Closed loop} (ki6);
      \draw[->] (ki5) -- (ki6);
      \draw[->] (ki6) -- (ki7);
      \draw[->] (ki7) -- (ki8);
      \draw[->] (ki8.west) -- ($(ki8.west)+(-4.5,0.0)$) |- node [near start] {No, Slave loop} (ki3);
      \draw[->] (ki8.south) -- node [near start] {Yes} (ki9);
      \draw[->] (ki8) -- (ki9);
      \draw[->] (ki9) -- (ki10);
    \end{tikzpicture}
    \caption{Flow chart of control software. \label{Fig:flowChart}}
\end{figure}
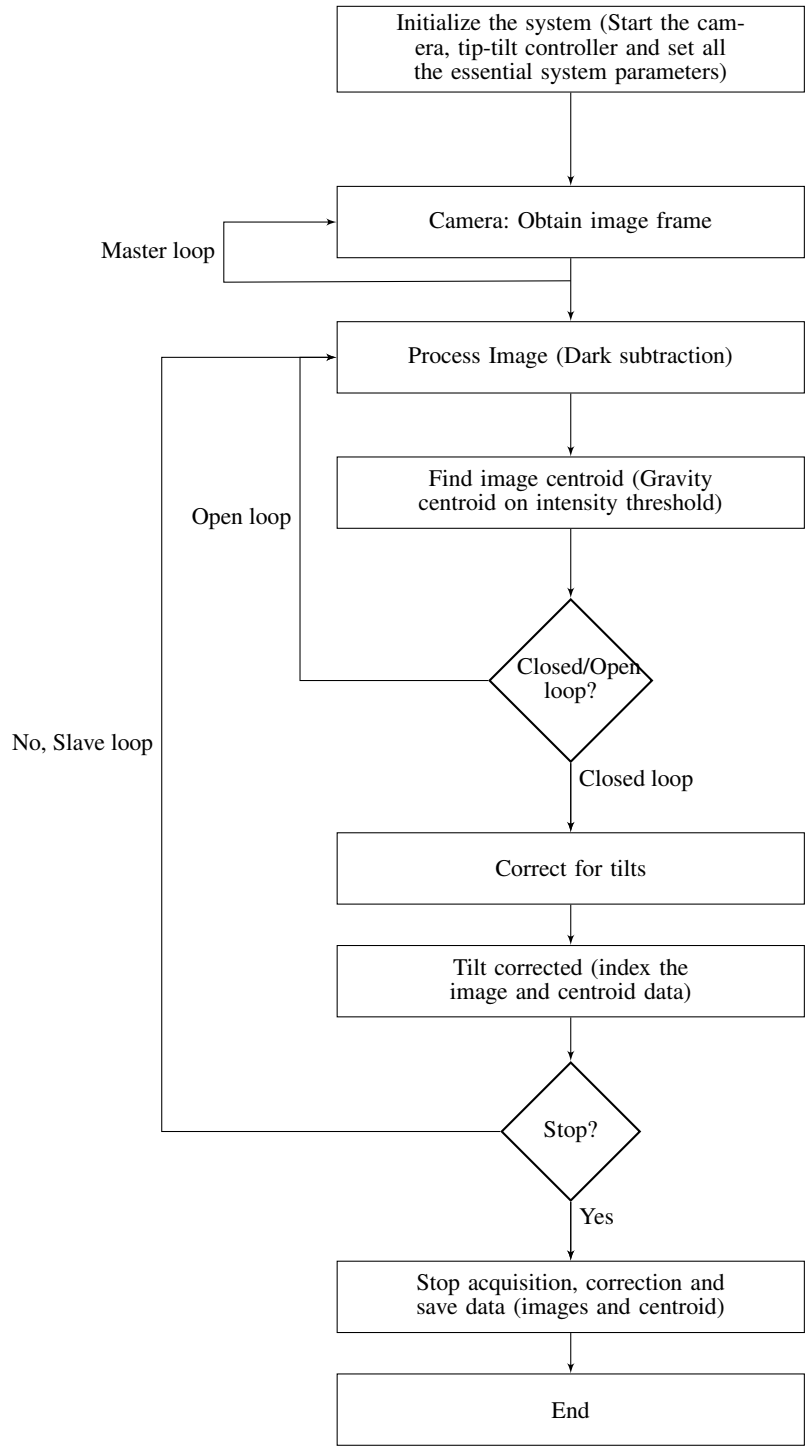

A control software with graphical user interface (GUI) is developed using National Instrument's LabVIEW platform\footnote{http://www.ni.com/} to operate the instrument. The software enables the interoperability between the sensing camera and the tip-tilt stage. The program flow of the control software is shown in the Figure \ref{Fig:flowChart}. As shown in the flowchart, during the initialization step,  the connectivity of the tip-tilt stage and the camera is checked. Subsequently, the  exposure time, the number of frames to obtain are set, following which the tip-tilt actuators are initialized. The program has one master loop and one slave loop where the former independently acquires the image frames, and the later processes them in sequence to estimate the centroid. The independence of the master loop from its slave improves the loop frequency of the system. The tip-tilt stage corrects for each of the centroid shifts in real time to compensate for the image motion. There is a time delay of $\approx$~0.8 ms, after acquiring the image and the tip-tilt mirror reaching its commanded position. The software has option to save the image and centroid data. For optimal performance of the instrument in closed loop, a PID controller was implemented in software. 

\begin{equation}
	\Delta c(t)=K_{p}e(t)+K_{d}\frac{de(t)}{dt}+K_{i}\int_{0}^{t^{\prime}} e(t^{\prime})dt
	\label {eq:eq4_1}\\
    \end{equation}
In the above equation, $K_{p}$, $K_{d}$ and $K_{i}$ are proportional, derivative and integral gains and $e(t)$ is the difference of two consecutive centroid shifts, $\Delta c(t)$ is the centroid shift estimated using the PID control. The controller gains are calculated using trial and error method by monitoring the residual image centroid motion. The estimated values for $K_{p}$, $K_{d}^{\prime}$ and $K_{i}^{\prime}$ are 4E-1, 2.5E-1, 2.3E-3 respectively. Here, $K_{d}^{\prime}$ and $K_{i}^{\prime}$ are $K_{d}*T$ and $K_{i}/T$, where $T$ is time interval between two consecutive centroids. The $\Delta c(t)$ is multiplied by voltage required to cause a pixel shift.
 
\subsection{Instrument calibration}
The calibration is required to accurately map the image wandering on the CCD to the input control voltage that drives the piezo actuators to compensate the image movement across the detector plane. This mapping is unique for each instrument as the linear beam-throw by steering mirror depends on the specific layout of the optical components. 

\begin{figure}
\centering
\includegraphics[height=4.5cm]{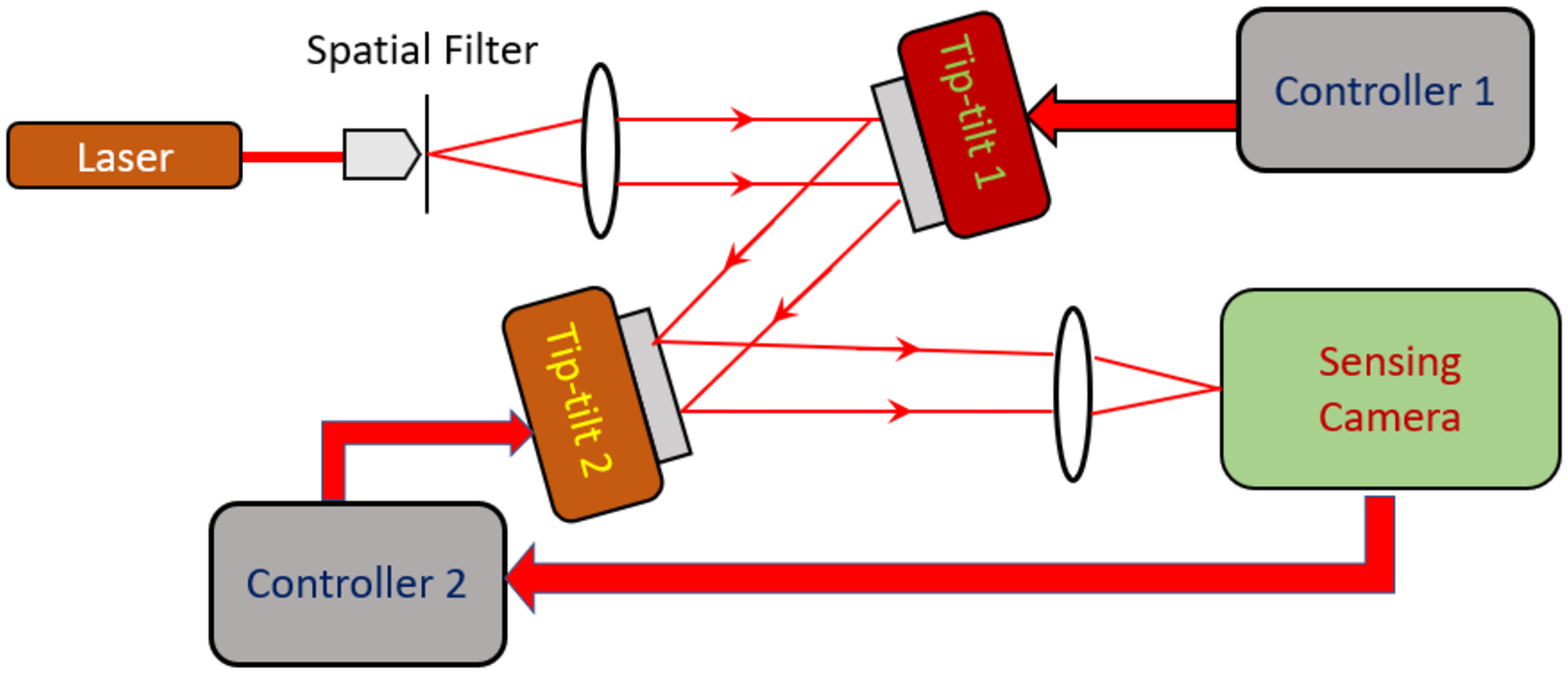}
\caption{ \label{fig:Labsetup} Experiment layout for the tip-tilt calibration in the lab. The tip-tilt stage TT1 was used to induce the image motion while TT2 was used to correct it. The CCD camera was used for recording the movement of the laser spot.}
\end{figure}

The axis of the stage have been centred around its maximum dynamic range i.e. at five mrad. The layout of laboratory setup is shown in the Figure \ref{fig:Labsetup}. A point source is generated by spatially filtering the laser beam as shown in the figure. The spatial filter consists of a microscope objective and a Pinhole of 10 $\mu$m. For performance analysis, the centroid motion of the point source image on sensing camera has been observed.

\subsubsection{Axis alignment}
Ideally, the movement of two actuator axis should be perfectly aligned with the pixel rows (horizontal) and columns (vertical) of the CCD. Initially, the axis of the tip-tilt stage were coarsely aligned by observing the spot traces in the live camera images. After making the fine adjustments,  a residual deviation in alignment  was measured by taking multiple images of the source on CCD while the beam was progressively steered (horizontally or vertically) in a sequence of voltage steps applied to individual actuators, one at a time. Furthermore, to minimize the alignment error, a rotation matrix (\citealt{russell71}) as shown in Equation \ref{eq:matrix} was generated from the obtained data. To compensate the small offset in the alignment the shift in the image motion is multiplied with this matrix before applying the corrections in real-time.  

\begin{equation}
\begin{bmatrix}
x^{\prime}\\
y^{\prime}\\
\end{bmatrix}
=
\begin{bmatrix}
a&b\\
c&d\\
\end{bmatrix}
*
\begin{bmatrix}
x\\
y\\
\end{bmatrix}
\label {eq:matrix}\\
\end{equation}

\begin{equation}
    X^{\prime}=AX
    \label {eq:rotation}\\
\end{equation}

\begin{equation}
    V=KX^{\prime}
    \label {eq:voltage}\\
\end{equation}

Equation \ref{eq:rotation} is the relation between the rotation matrix $A$ and the centroid shifts. In the Equation \ref{eq:matrix}, $x$ and $y$ are image centroid shift,  $x^{\prime}$ and $y^{\prime}$ are centroid shift after rotation along H-axis and V-axis. In our experiment, the measured elements of the rotation matrix ($A$) were: $a=1.003$,$b=0.0013$, $c=0.0015$ and $d=0.9996$. These are typical values for a closely aligned system.  In Equation \ref{eq:voltage}, $V$ is voltage applied to tip-tilt stage and $K$ is the voltage per unit shift in the centroid.

\begin{figure}
\centering
\includegraphics[height=9.5cm]{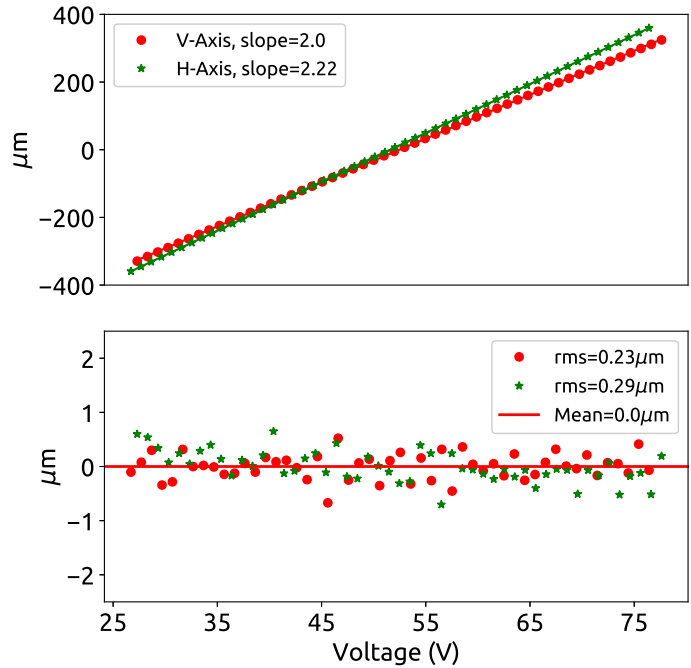}
\caption{ \label{fig:AxisCalibration} Calibration curves for two-axis tip-tilt stage. The horizontal (H-axis), vertical (V-axis) have been moved with equal step size of 1~V (top). In both cases, the rms deviation from the linear fit is $\sim$ 0.3~$\mu$m (bottom).}
\end{figure}

The response of each actuator is tested independently by tilting it over a range of 3~mrad to 8~mrad with an input voltage step size of 1V. The image centroids for each input step voltage is recorded. The mean centroid shift is estimated by measuring the average difference between two consecutive centroids over the earlier mentioned tilt range. Figure \ref{fig:AxisCalibration} (top panel) shows the image centroid shift on the camera as a function of input voltage. The error in actuator movement is defined as the difference between two consecutive centroids with respect to the mean centroid shift. It is plotted in the Figure \ref{fig:AxisCalibration} (bottom panel). The estimated \emph{rms} error in the actuator movement is less than 0.3 $\mu$m. Results of tip-tilt stage calibration are summarized in Table~\ref{tab:ttcalib}.

    \begin{table}
	\centering
	\caption{Tip-tilt stage axis calibration}
	\label{tab:ttcalib}
	\begin{tabular}{lccr} 
	\hline
	Parameter & Value\\
	\hline
	Number of samples & 51\\
	Exposure time per sample & 3 ms\\
	Tip-Tilt range & 3-8 mrad \\
	Voltage range & 24-78 V\\
	Voltage step size & $\sim$ 1V\\
	Linear image shift (total) & $\sim$650 $\mu$m\\
	Mean centroid shift & $\sim$ 12 $\mu$m\\
	rms error in movement & $\sim$ 0.3$\mu$m\\
	\hline
	\end{tabular}
\end{table}

\section{Laboratory testing of the instrument}\label{sec:labtest}
Initially, a prototype of the instrument was set up in the laboratory. Centroid data of short exposure images of the star was obtained a priory from the telescope. This data was then used as input to one of the steering mirrors to simulate the image motion in the lab studies. For characterization of the instrument, the residual image centroid motion and the PSD of tilt uncorrected and corrected images were analysed.

The layout of the laboratory setup is shown in Figure \ref{fig:Labsetup}. In this study we used two tip-tilt stages -one to induce the image motion (Piezosystems jena\footnote{$https://www.piezosystem.com/$}, Model Number: PSH x/2); and the other to correct it. The former has the frequency response up to 3~kHz, 0.02 $\mu$rad resolution and  $\pm$4 mrad dynamic range. 

The centroid data (in pixels) need to be converted to the voltages which will be applied to the tip-tilt stage 1 (TT1). For this purpose, the per-pixel voltage (0.21 V for H-axis and 0.23 V for V-axis) is estimated for the TT1 system from the calibration curves. The voltages were applied to TT1 to induce the image motion at the frequency of 33 Hz. This is to maintain at least 10 times correction bandwidth (\citealt{har98}) of the system. The induced image motion is tracked using the centroid estimations.

\begin{figure}
\centering
\includegraphics[height=7.5cm]{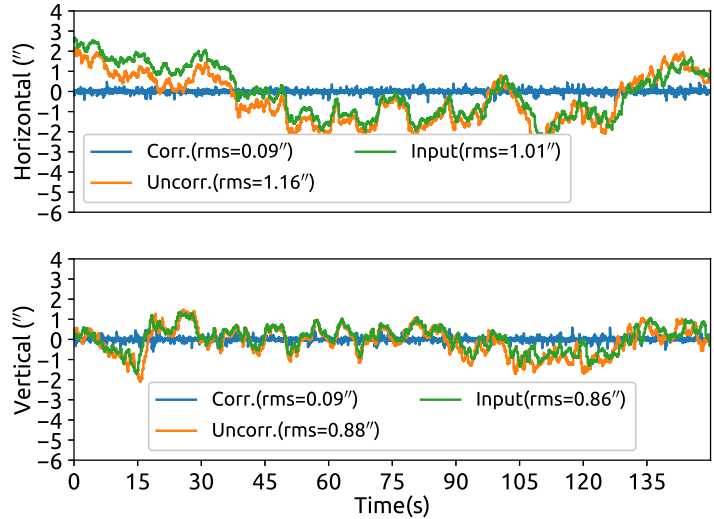}
\caption{ \label{fig:LabCentroid} Image centroid motion in arcsec ($^{\prime\prime}$) along horizontal (top panel) and vertical (bottom panel) axis of the camera. The plot has three sets of data i.e. the induced image motion, the uncorrected image motion and the corrected residual image motion. The induced and the sensed image motion has $\sim$ $96\%$ correlation. The image shift is converted to arc-seconds by multiplying the image shift with pixel scale similar to the telescope (0.06$^{\prime\prime}$).}
\end{figure}

The image motion data was recorded both with and without the tip-tilt correction. The Figure \ref{fig:LabCentroid} shows the image centroid motion in the laboratory. The \emph{rms} of corrected image motion is reduced by a factor of $\sim$ 12.8 in horizontal axis and $\sim$ 9.8 in vertical axis, compared to the uncorrected.

The closed loop correction bandwidth of the system was estimated  from the power spectral densities of the image motions for both tilt uncorrected and corrected cases. The observed correction bandwidth (0~dB)is $\sim 25 Hz$.

\begin{figure}
\centering
\includegraphics[height=7.5cm]{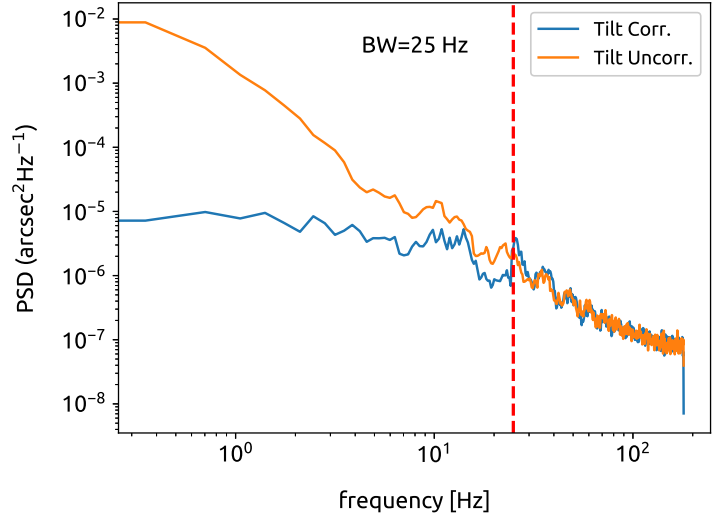}
\caption{ \label{fig:LabPSD} The power spectral density of centroid data in laboratory.  The vertical line signifies the merging of the two plots. This is the 0~dB closed loop correction bandwidth (red line) of the system.}
\end{figure}

\section{On-sky testing of the instrument}\label{sec:onsky}
After successful alignment and calibration, the instrument was mounted on the telescope (see Figure \ref{fig:instrumentMech}). Initial tests were done in March 2018. The on-sky performance of the tip-tilt system is described in terms of the residual \emph{rms}  image motion, correction bandwidth, full width half maximum  and peak intensity of the image. We are presenting these results after satisfactory performance has been achieved since January 2019. 

\subsection{Observations}
The objects of $m_v$ brighter than six have been chosen.  These objects were close to the zenith with hour angle of less than one hour. Preferably, targets with more than one object in the field have been chosen. This enables the instrument to sense the bright star with high-speed and apply the correction to the entire field. List of the targets used is given in the Table \ref{tab:Targets}. These targets were observed on different days from March 2019 to May 2019.

For high speed performance, a region of interest (ROI) has been chosen around the target image on sensing camera. This  enhanced the frame rate and thus increased the overall loop frequency. The exposure times are chosen from 3 ms to 20 ms. The longer exposure time allows us to observe the relatively fainter targets. This enables us to vary a loop frequency from $\sim$ 290 fps to 47 fps. 
Each data set was recorded over 150 seconds. Under poor atmospheric conditions, the binning operation was carried out in the program to enhance the signal to noise ratio. This might have reduced the accuracy in centroid estimation, but overall, it improved the correction performance.

\begin{figure}
\centering
\includegraphics[height=5.0cm]{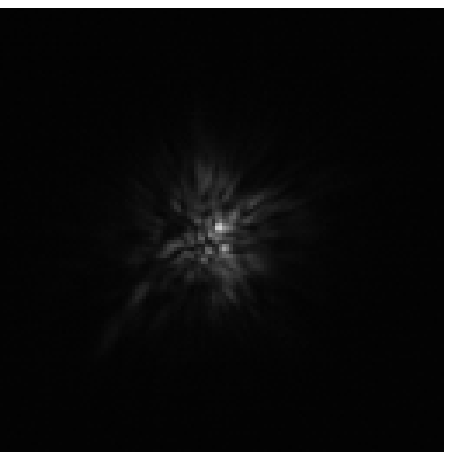}
\includegraphics[height=5.0cm]{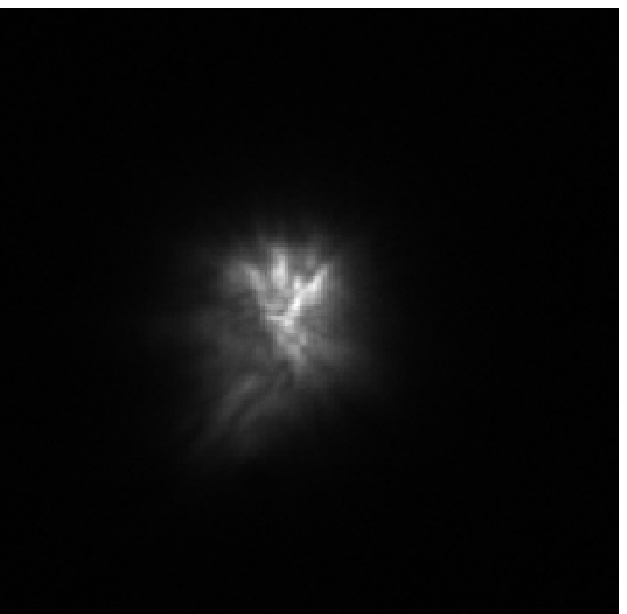}\\
\caption{ \label{fig:ShortExp} Illustration object: HIP57632. The short exposure time (3 ms) image on sensing camera (left). Relatively long exposure time (200 ms) image on the imaging camera. The frame size is $\sim$ $10^{\prime\prime}\times10^{\prime\prime}$.}
\end{figure}

The tilt corrected images were acquired on imaging camera. This camera was given relatively longer exposure time($\sim$200 ms). Because of a set of objects with different magnitude, we chose a fixed exposure time to avoid pixel saturation in case of a brighter object. In  Figure \ref{fig:ShortExp}, the short exposure image of sensing camera and the imaging camera were shown.

\begin{figure}
\centering
\includegraphics[height=13.5cm]{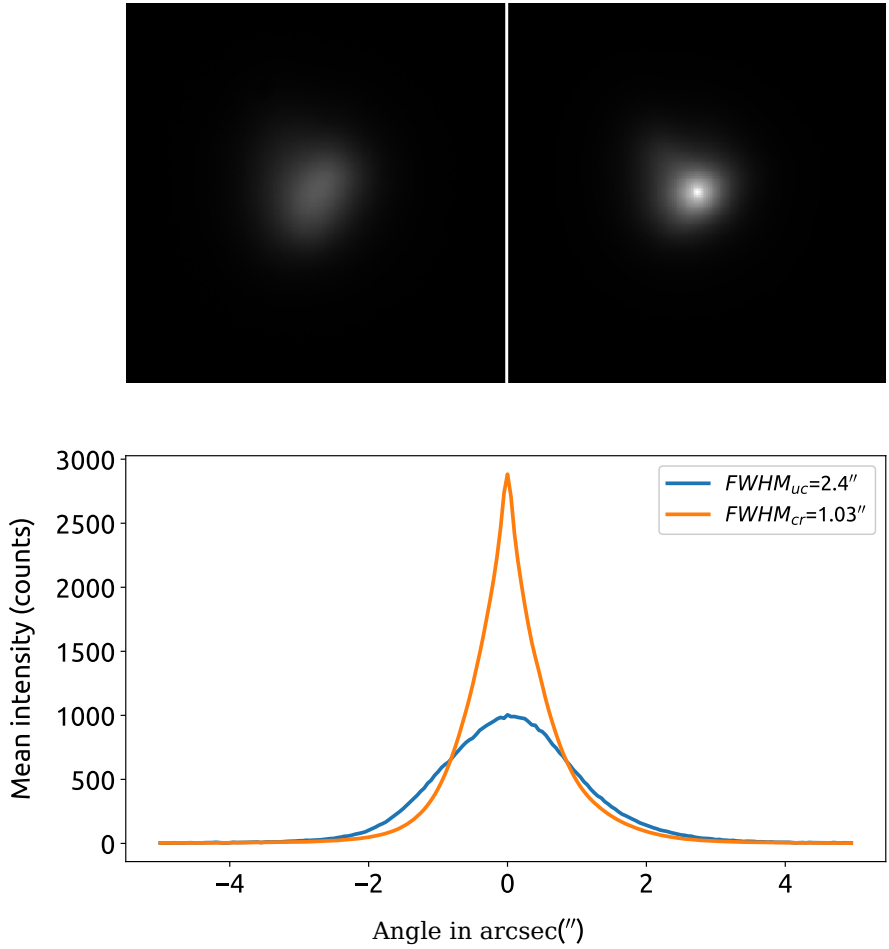}
\caption{ \label{fig:LongExp} Illustration: Object HIP57632. The tilt uncorrected (top-left) and the tilt corrected (top-right) co-added images. The psf of the image along the horizontal axis is shown in the bottom panel.}
\end{figure}

A total of 1000 images of each target field were acquired on the imaging camera. Every target was observed for tilt uncorrected and corrected images. These images were processed using a Python script. The image frames have been dark subtracted and flat fielded prior to the analysis.  The co-added images, as shown in Figure \ref{fig:LongExp}, will give the equivalently long exposure images. Finally, these images were divided by  number of obtained frames. This will average the intensity of each frame and minimize the effect of intensity fluctuations on the estimation of the performance.

\subsection{Image centroid, PSD and psf comparison}
\begin{figure}
\centering
\includegraphics[height=7.5cm]{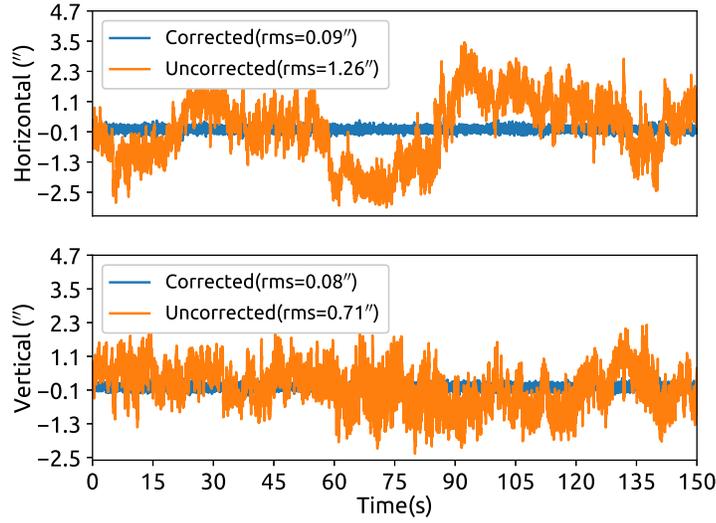}
\caption{ \label{fig:CentroidMotion} Image centroid motion of HIP57632. The rms image motion has been reduced by a factor of $\sim$ 14 in horizontal axis and $\sim$ 8.9 in vertical axis.}
\end{figure}

 The instrument was characterized by measuring the residual image motion. Figure \ref{fig:CentroidMotion} shows the tilt-corrected and uncorrected image centroids recorded consecutively by sensing camera. 
 The rms value was reduced to $\sim$0.08$^{\prime\prime}$ from $\sim$1.26$^{\prime\prime}$.

\begin{figure}
\centering
\includegraphics[height=7.5cm]{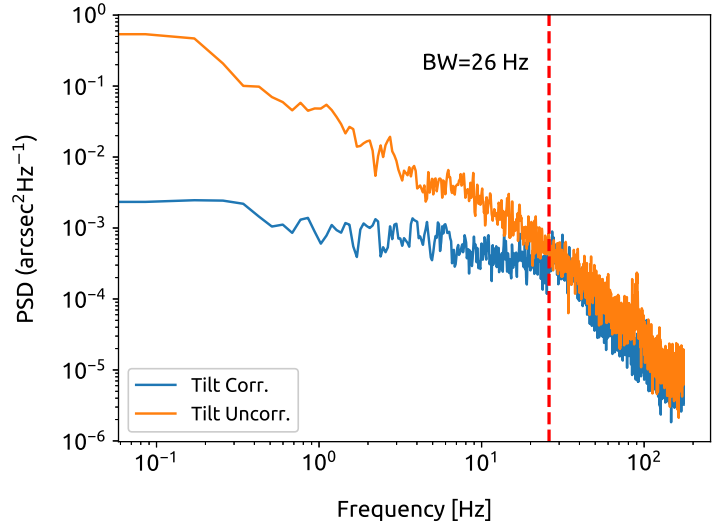}
\caption{ \label{fig:TelescopePSD} The power spectral density of the image motion for star HIP57632. Vertical dotted line demarcates the close-loop bandwidth of the system.}
\end{figure}

The image motion power spectral density is shown in Figure \ref{fig:TelescopePSD}. The correction bandwidth of the system is found to be  $\sim$26 Hz, depicted with a vertical line in the figure. The temporal frequencies beyond this limit are uncorrected.

On imaging camera, the psf is expected to be sharper with tip-tilt instrument in operation. The Figure \ref{fig:LongExp} shows the psf of HIP57632. The full width at half maximum (FWHM) of the psf has improved from 2.4$^{\prime\prime}$ to 1.03$^{\prime\prime}$.  The improvement in FWHM was calculated as shown in Equation \ref{eq:psf}. This is a $\sim$ 57\% improvement. 

\begin{equation}
    Improvement(\%)=\frac{FWHM_{uc}-FWHM_{cr}}{FWHM_{uc}}\times100
	\label {eq:psf}\\
    \end{equation} 
Here, $FWHM_{uc}$ is for uncorrected image and $FHWM_{cr}$ is for corrected image.

In the above case, the peak intensity of tilt corrected psf has increased by factor of $\sim$ 2.8. This improves the sensitivity of the instrument towards the observation of a fainter object. The sensitivity was estimated by using the Equation \ref{eq:eq6}. This is equation relates the magnitude difference of a star with peak intensity of the tilt uncorrected and corrected images. It is observed that the sensitivity is improved by a factor of 1.1 in magnitude.

\begin{equation}
    \Delta m_{v}=-2.512*log_{10}\Big[\frac{I_{c}}{I_{uc}}\Big]
	\label {eq:eq6}\\
    \end{equation}    
    Where, ${I_{c}}/{I_{uc}}$ is the ratio of peak intensities of tilt corrected and uncorrected images, $\Delta m_{v}$ is the improvement in apparent magnitude. Here, we considered, ${I_{c}}/{I_{uc}}$ as ${4.3}/{1.49}$.

\subsection{Effect of loop frequency}
The optimal frame rate is essential for the effective tilt correction of the images in a close-loop operation. This is in confirmation with the fact that the wave-front distortions are caused by the spatial and the temporal disturbances in the atmosphere. The spatial distortions are corrected by compensating for the shift in the image centroid. But the dynamic nature of the atmosphere induces high frequency image motion. To overcome this effect, the time delay between the instant of estimation of the shifts and the instant of the correction applied to the corrector should be kept minimum. 

To study the effect of loop frequency  on the peak intensity of the tilt-corrected images, we observed HIP57632 with different frame rates. The frame rate was changed by changing the exposure time of the sensing camera from three to 20 ms, yielding loop frequencies of 290 to 47 Hz.

\begin{figure}
\centering
\includegraphics[height=8.5cm]{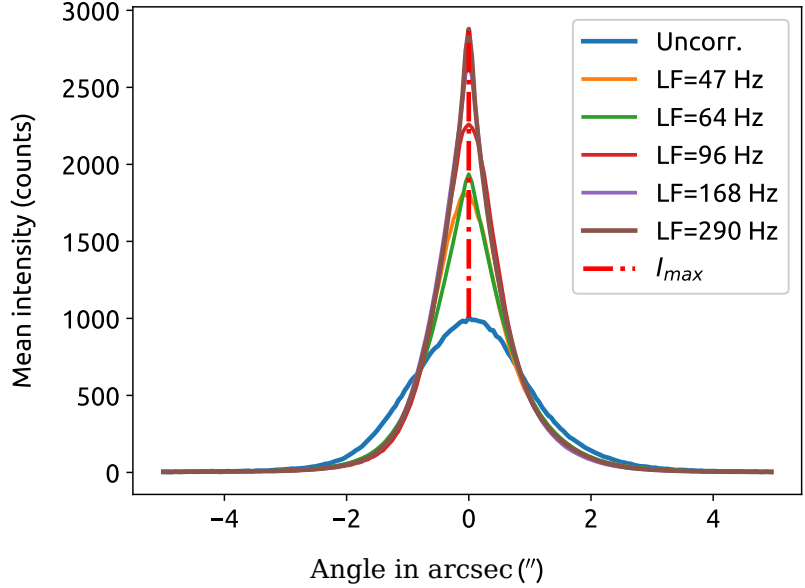}
\caption{ \label{fig:psfcomp} Comparison of tip-tilt corrected psf of HIP57632 for different loop frequencies. The peak intensity is showing logistic growth with loop frequency.}
\end{figure}

\begin{figure}
\centering
\includegraphics[height=8.5cm]{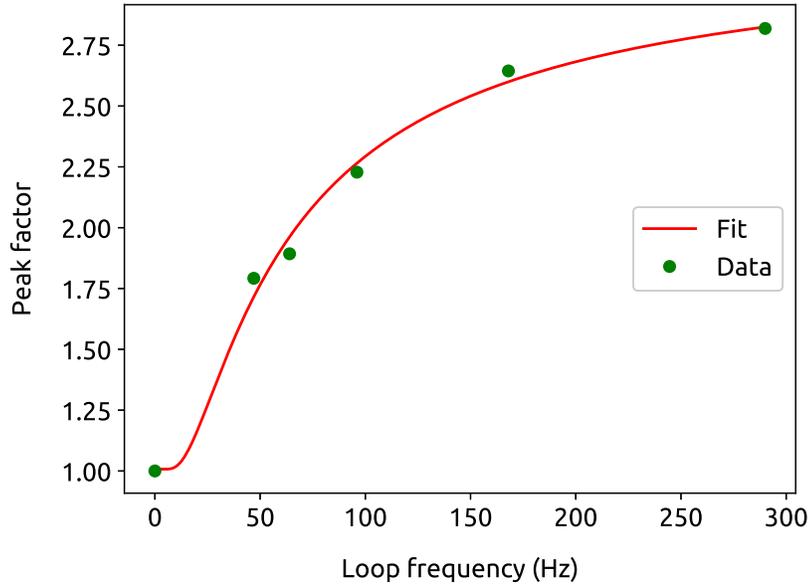}
\caption{ \label{fig:psflf} Increment in peak factor with loop frequency.}
\end{figure}

In Figure \ref{fig:psfcomp}, cross section of the psfs with different loop frequency are plotted.  The peak intensity of the psf was increased with increase in loop frequency and the result is plotted in Figure \ref{fig:psflf}. In this figure, the peak factor was defined as the ratio of peak intensity of the tilt corrected to that of uncorrected image. 

We have modeled the effect of loop frequency on peak factor with a function of the form shown in Equation \ref {eq:eq5}.
\begin{equation}
    I_{pf}(f)=K_1+K_2*(1-e^{-f_0/f})
	\label {eq:eq5}\\
    \end{equation} 
In the above equation, $f$ is loop frequency, $I_{pf}$ is peak factor and the estimated constants are $K_{1}$, $K_{2}$ and $f_0$ estimated to be $\approx$ 3.2, -2.2 and 52 Hz, respectively. The units of $K_2$ and $K_1$ are similar to $I_{max}$. Arguably, the estimated values of the constants depend on the target intensity and the atmospheric seeing conditions.  

\begin{figure}
\centering
\includegraphics[height=9.5cm]{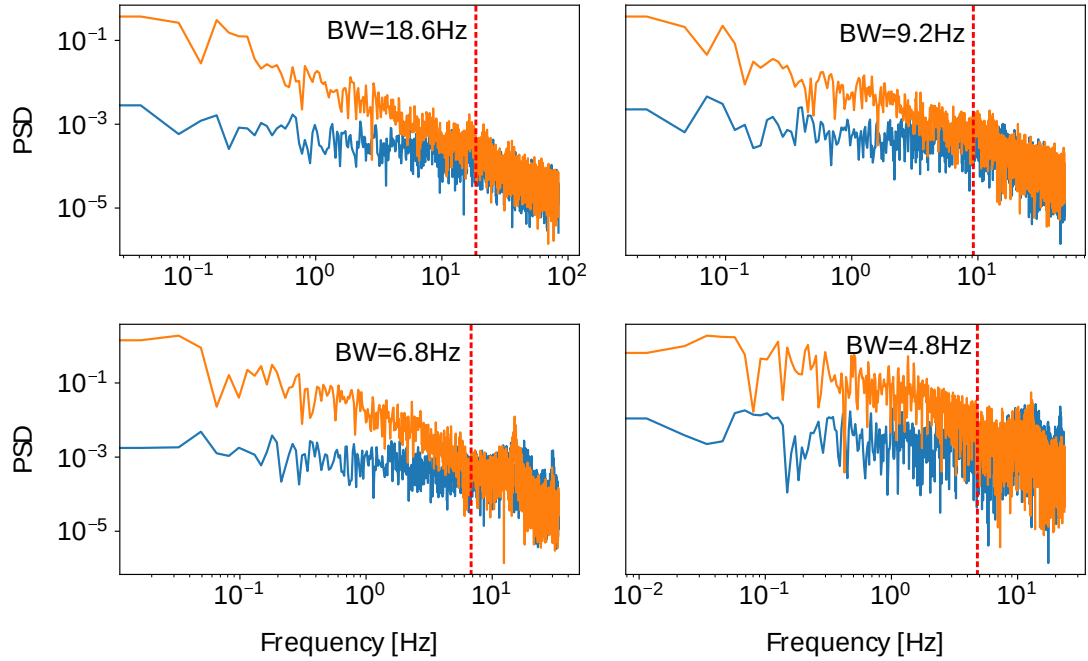}
\caption{ \label{fig:psdlf} The PSD (arcsec$^2$Hz$^{-1}$) with respect to correction bandwidth for object HIP57632. The orange and the blue lines are PSD of tilt uncorrected and corrected centroid data. The loop frequency is 168 Hz (left) and 96 Hz (right) for top row, 64 Hz (left) and 47 Hz (right) for bottom row and their correction bandwidths (red line) are in the plot.}
\end{figure}

In Figure \ref{fig:psdlf}, the power spectral densities for different loop frequencies are shown. We can see that that the correction bandwidth increased to $\sim$ 26 Hz at  290 Hz  from $\sim$ 4.8 Hz at 47 Hz loop frequency.  On an average, the correction bandwidth is $\sim$ 1/10 of the loop frequency which is in agreement with other studies reported in literature (\citealt{har98}). 

\subsection{Gain in angular resolution}
\begin{figure}
\centering
\includegraphics[height=7.5cm]{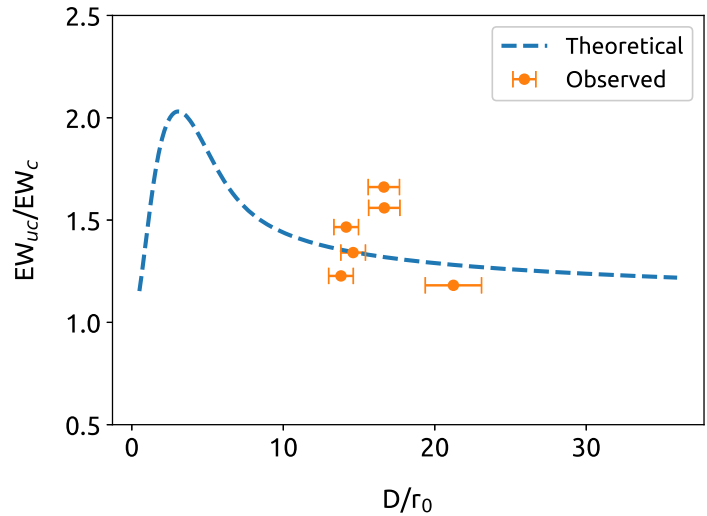}
\caption{ \label{fig:angain} Comparison of gain in angular resolution between the theoretical and the observed data.  On vertical axis, EW$_{uc}$ and EW$_{c}$ are equivalent widths of tilt uncorrected and corrected image.}
\end{figure}

The tilt corrected images show improvement in angular resolution. The gain in angular resolution is a function of  relative sizes of  the telescope aperture (D) and the atmospheric coherence diameter $r_{0}$. This relation can be theoretically estimated using the formalism given by \cite{roddier1981v} and is shown in Figure \ref{fig:angain}.  The gain is defined as the ratio of equivalent width (\citealt{roddier1981v}) of tilt uncorrected image and corrected images.  The observed values of the gain for a set of six targets as listed in Table \ref{tab:Targets}  are over plotted on the theoretical curve. $r_0$ is estimated from the equivalent width of the uncorrected images (to get the $D/r_0$ for the observed data). The observed gain has shown a deviation up to 24\% from the theoretical gain.

\subsection{Performance of the instrument on faint targets}
We observed a set of seven objects as listed in Table \ref{tab:Targets} (objects 7-13) to validate the increase in sensitivity of the instrument due to image stabilization. Usually, a bright star near faint star is used for sensing the image motions and the same correction is applied to the entire field.  If the faint star is close enough, the corrections are similar and thus the sensitivity of the instrument on the faint star increases.

\begin{figure}
\centering
\includegraphics[height=7.5cm]{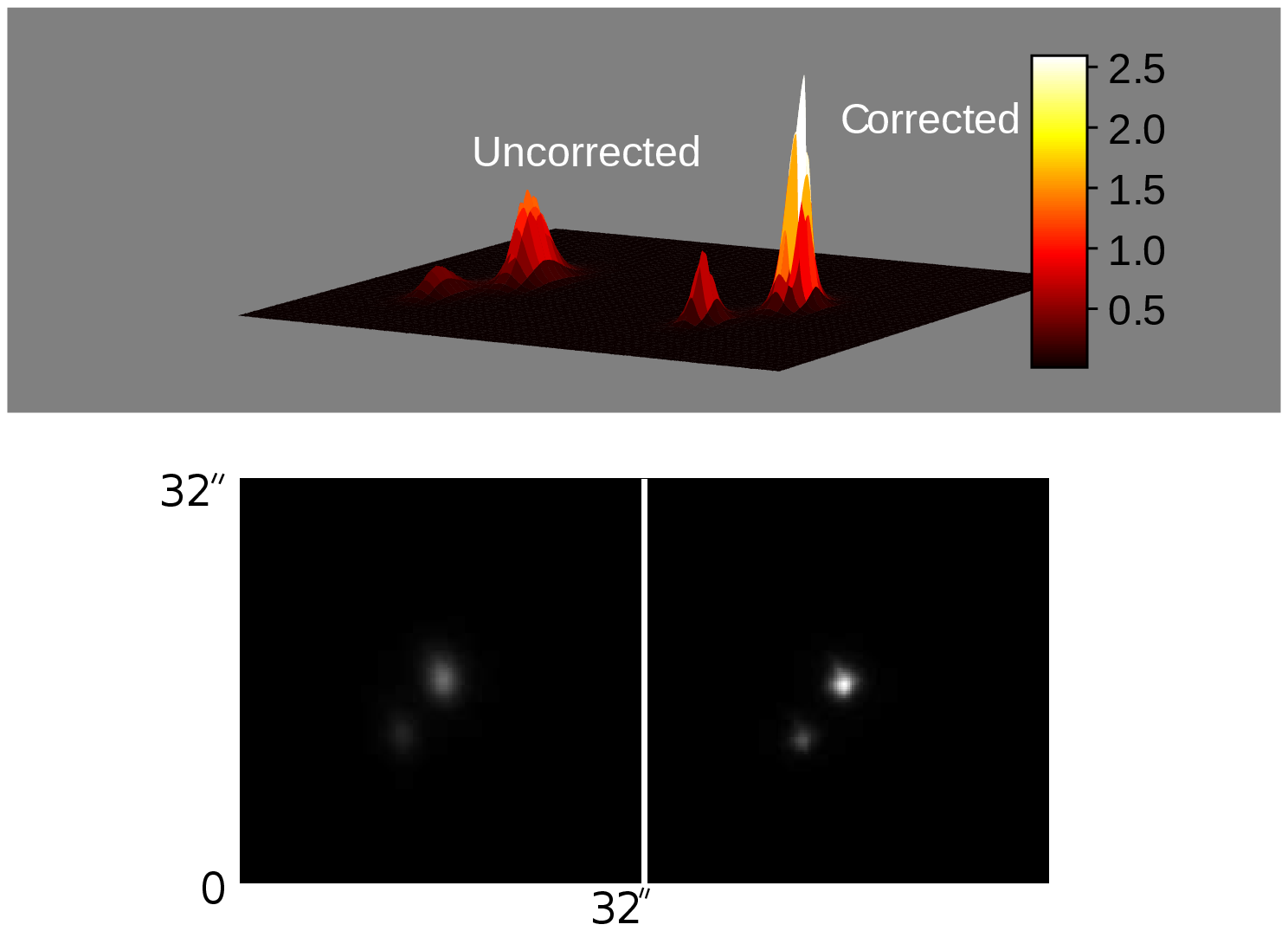}
\caption{ \label{fig:PSFHIP50583} Surface plot of HIP50583, containing  average tilt uncorrected and corrected image. The frame size is $\sim$ $32^{\prime\prime}\times32^{\prime\prime}$.}
\end{figure}

\begin{figure}
\centering
\includegraphics[height=8.5cm]{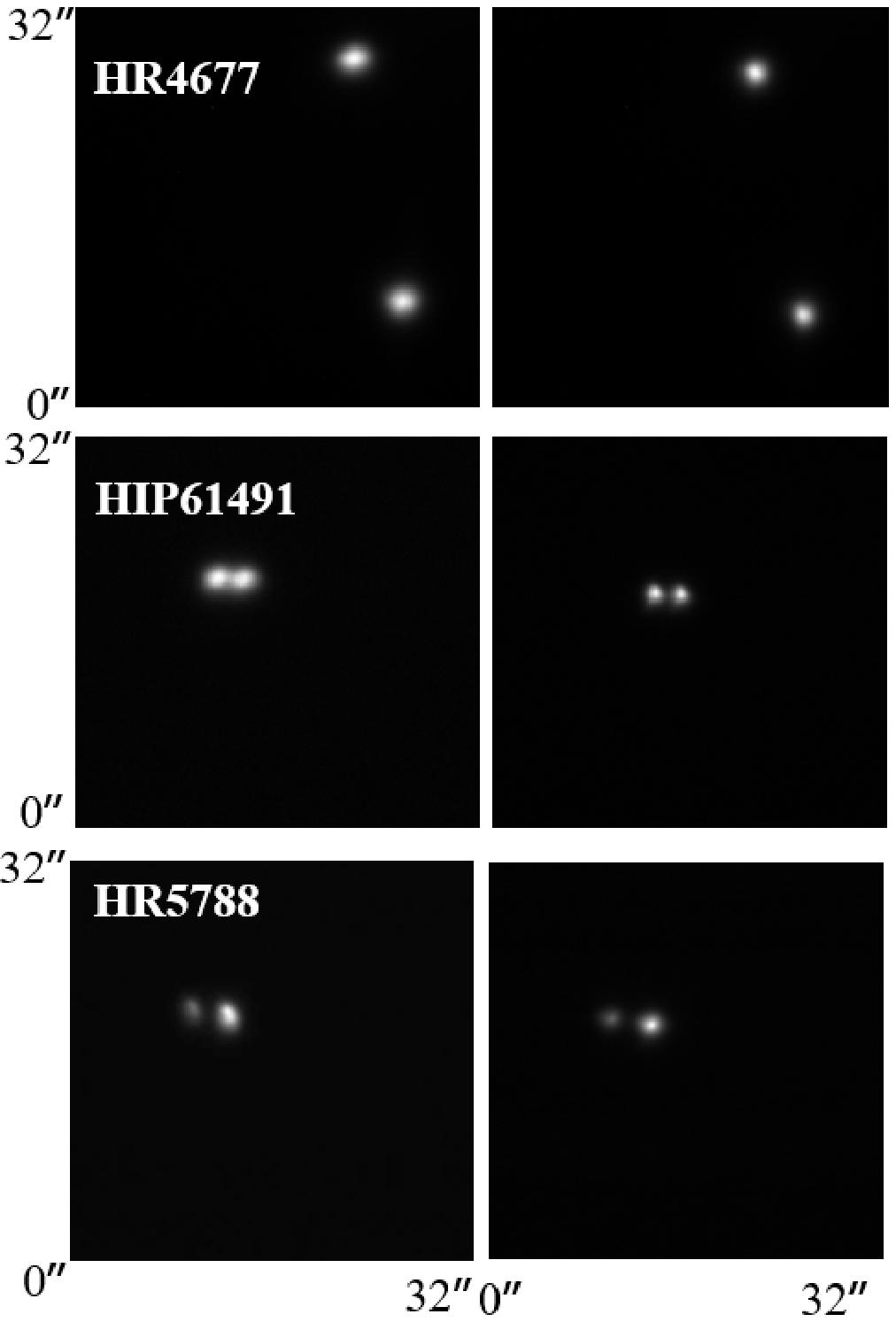}
\includegraphics[height=8.5cm]{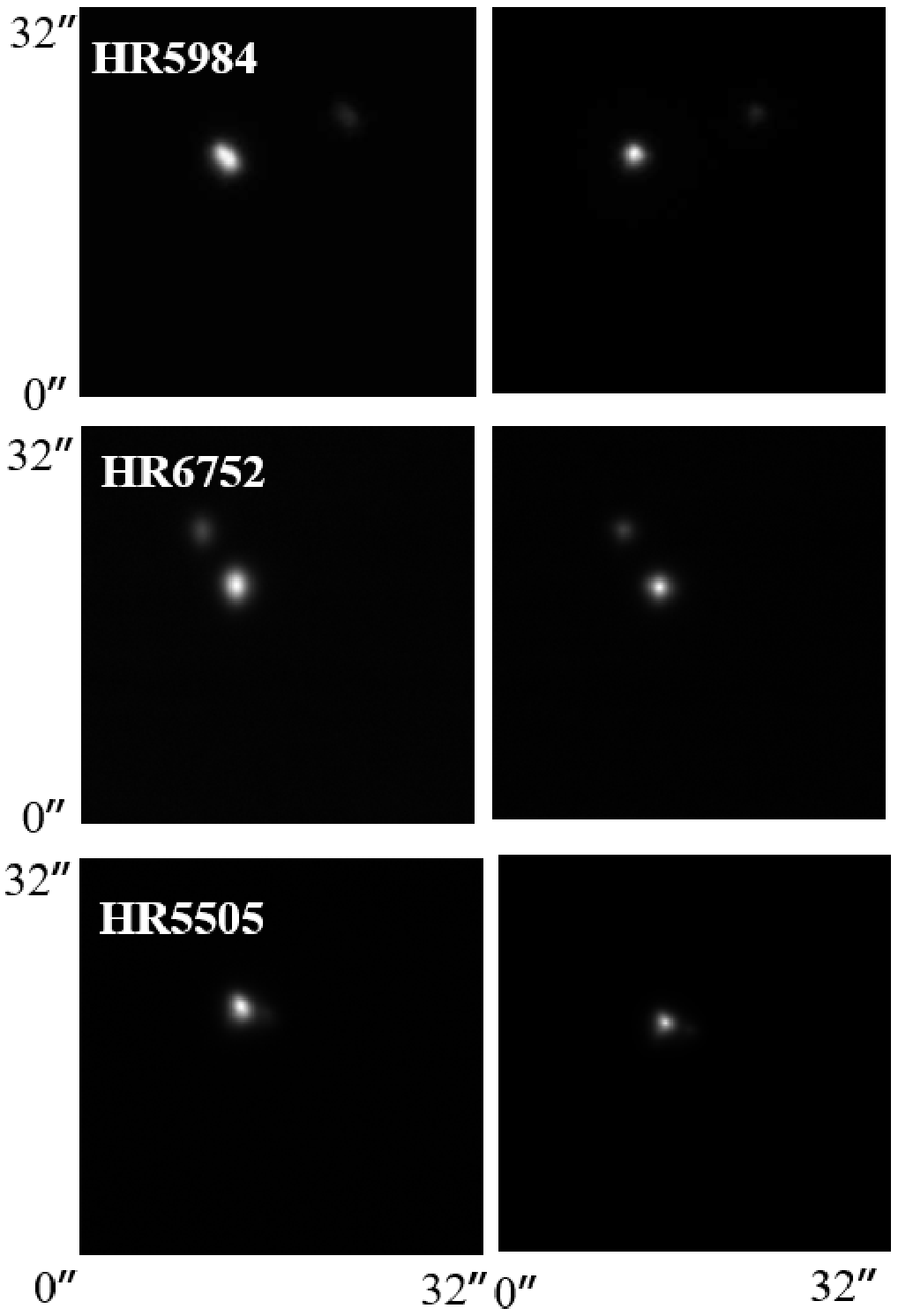}\\
\caption{ \label{fig:twostars} Examples: Tilt uncorrected and corrected images of various targets.}
\end{figure}

In Figure \ref{fig:PSFHIP50583} the tilt uncorrected and corrected image of  HIP50583 is shown. The object has brightness of 2.37 in magnitude with a relatively fainter object with magnitude of 3.47, at an angular separation of 4.63$^{\prime\prime}$. The correction increased the peak intensity by a factor of  $\sim$ 2.5 times in brighter object $\sim$ 2.1 times in fainter object. The angular resolution (FWHM) of these objects improved by 52\% and 46\% respectively. Similar to this, several other objects were observed and these results are shown in Figure \ref{fig:twostars} and Table \ref{tab:Targets}. 

\begin{table}
	\centering
	\caption{List of observed stars with varying angular separation. In the table, $m_v$ is apparent magnitude, $\Delta m_v$ is magnitude different between two objects and 'Sep' is angular separation between the objects in arc-sec. Peak factor is the ratio of tilt corrected and uncorrected image and improvement in resolution ($R$) is similar to Equation \ref{eq:psf}. LF is approximate loop frequency}
	\label{tab:Targets}
	\begin{tabular}{lcccccccccr} 
	      \hline
	      Sl.No.&Target&RA&Dec&$m_v$& $\Delta m_v$& Sep(${\prime\prime}$)& LF (fps)& Peak factor& R(\%) \\
	      \hline
	      1&HIP57632,$\--$ &11 49 03.5&+14 34 19.4&2.13& $\--$& $\--$&290&2.8,$\--$&57,$\--$ \\
	      2&HIP54879,$\--$&11 14 14.4& +15 25 46.4&3.35& $\--$& $\--$&290&2.1,$\--$&45,$\--$ \\
	      3&HIP37279,$\--$&07 39 18.1&+05 13 29.9&0.37& $\--$& $\--$&290&1.8,$\--$&44,$\--$ \\
	      4&HIP65474,$\--$&13 25 11.5&-11 09 40.7&0.97& $\--$& $\--$&290&1.4,$\--$&32,$\--$ \\
	      5&HIP67927,$\--$&13 54 41.0&+18 23 51.7&2.68&$\--$& $\--$&290&1.5,$\--$&38,$\--$ \\
	      6&HIP69673,$\--$&14 15 39.6&+19 10 56.6&-0.05&$\--$& $\--$&290&2.4,$\--$&50,$\--$ \\
	      7&HIP50583, gam02 Leo&10 19 58 &19 50 29.3&2.37&1.1&4.63&290& 2.5, 2.1&52, 46\\
	      8&HIP61941, gam Vir B& 12 41 39.6& -01 26 57.7&2.74&0.75&1.52&290&2.4, 2.3&51, 49\\
	      9&HR4677,  	HD 106976&12 18 08&-03 57 05.01&5.99&0.7&20.33&64&1.5, 1.3&36, 32\\
	      10&HR6752, 70 Oph B&18 05 27 & 02 30 0.0&4.03 &2.04&4.91 &98&1.9, 1.7&41,36 \\
	      11&HR5789, del Ser B&15 34 48.1& 10 32 15.9 &3.79 &1.4 &4.1&98&1.7, 1.5 &32, 33\\
	      12&HR5984, bet02 Sco&16 05 26.2 & -19 48 19.6& 2.5&2.3&13.64&98&1.8, 1.6&42, 36\\
  	      13&HR5505, eps Boo B& 14 44 59.2 &+27 04 27.2&2.39&2.4&2.58&98&1.8, 1.5&43, 31\\
	      \hline
	  \end{tabular}
\end{table}

\section{Summary and conclusion}\label{sec:discuss}
 A tip-tilt instrument has been developed for the 1.3 m JCB telescope to overcome the image degradation caused by angle of arrival fluctuations. In laboratory, a simulated image motion corresponding the actual data acquired from the telescope was used to characterize the instrument. The on-sky performance of the instrument was analyzed by observing  several stars of varying brightness and angular separation. The real-time correction has shown a characteristic improvement in the image quality that is consistent with  previously reported studies in the literature. This study has led to the following findings and conclusions: 
\begin{enumerate}
    \item In laboratory, the rms image motion was reduced by $\sim$12 times and the correction bandwidth was estimated to be a $\sim$ 25 Hz for a loop frequency of 290 Hz.
    \item On telescope, the rms image motion was reduced by $\sim$14 times and the correction bandwidth was estimated to be about 0.1 times the loop frequency, where the loop frequency was varied from 47 Hz to 290 Hz (five distinct frequencies in this range).
    \item The FWHM of the image reduced from 2.4$^{\prime\prime}$ to 1.03$^{\prime\prime}$. This corresponds to 57\% improvement in the image resolution.
    \item The sensitivity of the instrument was found to increased by a factor of 1.1 magnitude (corresponding to the increase in the dynamic range, peak intensity ratio of 2.8).

\item In the case of targets with two close-by stars in the field, the FWHM of the individual psf decreased and the peak brightness increased depending on the magnitudes. For example, in the case of  HIP50583 with separation of 4.63$^{\prime\prime}$ and magnitude difference of 1.1,  the FWHM of the bright star increased by 52\% and that of the faint star increased by 46\%.  The peak brightness of the bright star increased by a factor of $\sim$ 2.5  and that of the faint star increased by a factor of $\sim$ 2.1.
  \end{enumerate}
  
To further improve the image quality to near diffraction limited resolution of the telescope, the work on a higher order AO system is under progress. 

\section*{Acknowledgements}
We would like to thank Prof. B Raghavendra Prasad for allowing us to use the lab facilities at CREST. We also gratefully acknowledge the help received from Mr Suresh Venkata Narra during the initial testing and Mr Hari Mohan Varsey for preparing with the mechanical design of the instrument. We thank Mr Anbazhagan, the engineer-in-charge, VBO Kavalur, for providing logistics and technical support during the on-sky testing. We also thank the observing staff at VBO, namely, V. Moorthy, G. Selva Kumar, S Venkatesh, Rahul and Naveen for their valuable support during this work.
  
\bibliographystyle{raa}
\bibliography{ms2019-0208bibtex}

\begin{thebibliography}{25}
\providecommand\natexlab[1]{#1}
\providecommand\JournalTitle[1]{#1}

\bibitem[Close \& McCarthy(1994)]{close94}
Close, L., \& McCarthy, D. 1994, Publications of the Astronomical Society of
  the Pacific, 106, 77

\bibitem[Dainty {et~al.}(1998)]{dainty98}
Dainty, J.~C., Koryabin, A.~V., \& Kudryashov, A.~V. 1998, Applied optics, 37,
  4663

\bibitem[{Davis} \& {Tango}(1996)]{davis96}
{Davis}, J., \& {Tango}, W. 1996, \pasp, 108, 456

\bibitem[Eaton {et~al.}(1985)]{eaton85}
Eaton, F., Peterson, W., Hines, J., \& Fernandez, G. 1985, Applied optics, 24,
  3264

\bibitem[{Fried}(1965)]{fri65}
{Fried}, D.~L. 1965, J. Opt. Soc. Am. A, 55, 1427

\bibitem[{Fried}(1966{\natexlab{a}})]{Fried66}
{Fried}, D.~L. 1966{\natexlab{a}}, J. Opt. Soc. Am. A, 56, 1372

\bibitem[{Fried}(1966{\natexlab{b}})]{fri66}
{Fried}, D.~L. 1966{\natexlab{b}}, J. Opt. Soc. Am. A, 56, 1372

\bibitem[Glindemann(1997)]{glindemann97}
Glindemann, A. 1997, Publications of the Astronomical Society of the Pacific,
  109, 682

\bibitem[Glindemann {et~al.}(1997)]{glindemann1997charm}
Glindemann, A., McCaughrean, M.~J., Hippler, S., {et~al.} 1997, Publications of
  the Astronomical Society of the Pacific, 109, 688

\bibitem[Golimowski {et~al.}(1992)]{golimowski92}
Golimowski, D., Clampin, M., Durrance, S., \& Barkhouser, R. 1992, Applied
  optics, 31, 4405

\bibitem[Greenwood(1977)]{greenwood77}
Greenwood, D.~P. 1977, JOSA, 67, 390

\bibitem[{Hardy}(1998)]{har98}
{Hardy}, J.~W. 1998, {Adaptive Optics for Astronomical Telescopes}
  (Oxford.Univ.press), 448

\bibitem[Hubbard {et~al.}(1979)]{hubbard79}
Hubbard, G., Hege, K., Reed, M., {et~al.} 1979, The Astronomical Journal, 84,
  1437

\bibitem[{Kellerer} \& {Tokovinin}(2007)]{kel07}
{Kellerer}, A., \& {Tokovinin}, A. 2007, \aap, 461, 775

\bibitem[{Martin}(1987)]{mar87}
{Martin}, H.~M. 1987, \pasp, 99, 1360

\bibitem[Max {et~al.}(1997)]{max97}
Max, C.~E., Olivier, S.~S., Friedman, H.~W., {et~al.} 1997, Science, 277, 1649

\bibitem[Noll(1976)]{noll76}
Noll, R.~J. 1976, JOSA, 66, 207

\bibitem[Racine \& McClure(1989)]{racine89}
Racine, R., \& McClure, R.~D. 1989, Publications of the Astronomical Society of
  the Pacific, 101, 731

\bibitem[Roddier(1981)]{roddier1981v}
Roddier, F. 1981, in Progress in optics, Vol.~19 (Elsevier), 281

\bibitem[Rousset {et~al.}(1990)]{rousset90}
Rousset, G., Fontanella, J., Kern, P., Gigan, P., \& Rigaut, F. 1990, Astronomy
  and Astrophysics, 230, L29

\bibitem[Russell(1971)]{russell71}
Russell, C.~T. 1971, Cosmic Electrodynamics, 2, 184

\bibitem[Sarazin \& Tokovinin(2002)]{sarazin02}
Sarazin, M., \& Tokovinin, A. 2002, in European Southern Observatory Conference
  and Workshop Proceedings, Vol.~58, 321

\bibitem[Sreekanth {et~al.}(2019)]{VSReddy2019}
Sreekanth, R.~V., Banyal, R.~K., Sridharan, R., \& Selvaraj, A. 2019, Research
  in Astronomy and Astrophysics, 19, 074

\bibitem[Welch(1967)]{welch67}
Welch, P. 1967, IEEE Transactions on audio and electroacoustics, 15, 70

\bibitem[Wizinowich {et~al.}(2000)]{wizinowich2000}
Wizinowich, P.~L., Acton, D.~S., Lai, O., {et~al.} 2000, in Adaptive Optical
  Systems Technology, Vol. 4007, International Society for Optics and
  Photonics, 2

\end{thebibliography}
\end{document}